\documentclass[prl,twocolumn,superscriptaddress,amsmath,amssymb]{revtex4-1}
\usepackage{verbatim}
\usepackage{graphicx,float}
\usepackage{color}
\usepackage{bm}
\usepackage{dsfont}
\usepackage{epstopdf}
\usepackage{hyperref}
\usepackage{soul}

\begin{document}

\title{Correlating dynamic strain and photoluminescence of solid-state defects with stroboscopic x-ray diffraction microscopy}

\author{S. J. Whiteley}
\affiliation{Pritzker School of Molecular Engineering, University of Chicago, Chicago, Illinois 60637, USA}
\affiliation{Department of Physics, University of Chicago, Chicago, Illinois 60637, USA}
\author{F. J. Heremans}
\affiliation{Pritzker School of Molecular Engineering, University of Chicago, Chicago, Illinois 60637, USA}
\affiliation{Center for Molecular Engineering and Materials Science Division, Argonne National Laboratory, Lemont, Illinois 60439, USA}
\author{G. Wolfowicz}
\affiliation{Pritzker School of Molecular Engineering, University of Chicago, Chicago, Illinois 60637, USA}
\affiliation{Center for Molecular Engineering and Materials Science Division, Argonne National Laboratory, Lemont, Illinois 60439, USA}
\affiliation{WPI-Advanced Institute for Materials Research (WPI-AIMR), Tohoku University, Japan}
\author{D. D. Awschalom}
\affiliation{Pritzker School of Molecular Engineering, University of Chicago, Chicago, Illinois 60637, USA}
\affiliation{Center for Molecular Engineering and Materials Science Division, Argonne National Laboratory, Lemont, Illinois 60439, USA}
\author{M. V. Holt}
\email{mvholt@anl.gov}
\affiliation{Center for Nanoscale Materials, Argonne National Laboratory, Lemont, Illinois 60439, USA}

\begin{abstract}
Control of local lattice perturbations near optically-active defects in semiconductors is a key step to harnessing the potential of solid-state qubits for quantum information science and nanoscale sensing.  We report the development of a stroboscopic scanning X-ray diffraction microscopy approach for real-space imaging of dynamic strain used in correlation with microscopic photoluminescence measurements.  We demonstrate this technique in 4H-SiC, which hosts long-lifetime room temperature vacancy spin defects.  Using nano-focused X-ray photon pulses synchronized to a surface acoustic wave launcher, we achieve an effective time resolution of $\sim$100~ps at a 25~nm spatial resolution to map micro-radian dynamic lattice curvatures.  The acoustically induced lattice distortions near an engineered scattering structure are correlated with enhanced photoluminescence responses of optically-active SiC quantum defects driven by local piezoelectric effects.  These results demonstrate a unique route for directly imaging local strain in nanomechanical structures and quantifying dynamic structure-function relationships in materials under realistic operating conditions. 
\end{abstract}

\date{\today}
\maketitle

Strain provides a fundamental route to  control diverse material properties such as electrical transport \cite{Wu2016}, chemical reactivity \cite{Oakes2016}, and electromagnetic ordering \cite{Sando2013}.  In quantum systems, the manipulation of strain near isolated point defects and engineered structures has shown the potential to significantly improve performance characteristics of solid-state qubits for quantum information processing \cite{Ladd2010,Gustafsson2014,Heremans2016,Bassett2011,delasCasas2017}.  Among potential degrees of freedom in solid-state quantum devices, mechanics has the  ability to nearly universally interact with  various types of qubits \cite{Kurizki2015,Schuetz2015} and be confined to sub-micron length scales. Near-defect lattice strain can be used either statically, to tune quantum energy levels and degeneracy, or dynamically, to mechanically drive coherent spin transitions \cite{Barfuss2015,Macquarrie2015,Whiteley2018} and engineer hybrid system responses.  From this, mechanical systems have the potential to play a transformative role in quantum information transfer, but the degree and nature of strain coupling to local properties, such as defect electronic energy levels are often not well understood \cite{Weber2010,Falk2014} due to the difficulty of directly measuring local nanoscale strain congruently to optical response.  Quantifying this coupling is especially important in the time domain, where dynamic sources of strain such as resonant acoustic waves can be used to manipulate spin states or control the transmission of single electron currents between qubits \cite{Schulein2015,Hermelin2011}.

\begin{figure*}[t]
\centerline{\includegraphics[width=6.95in]{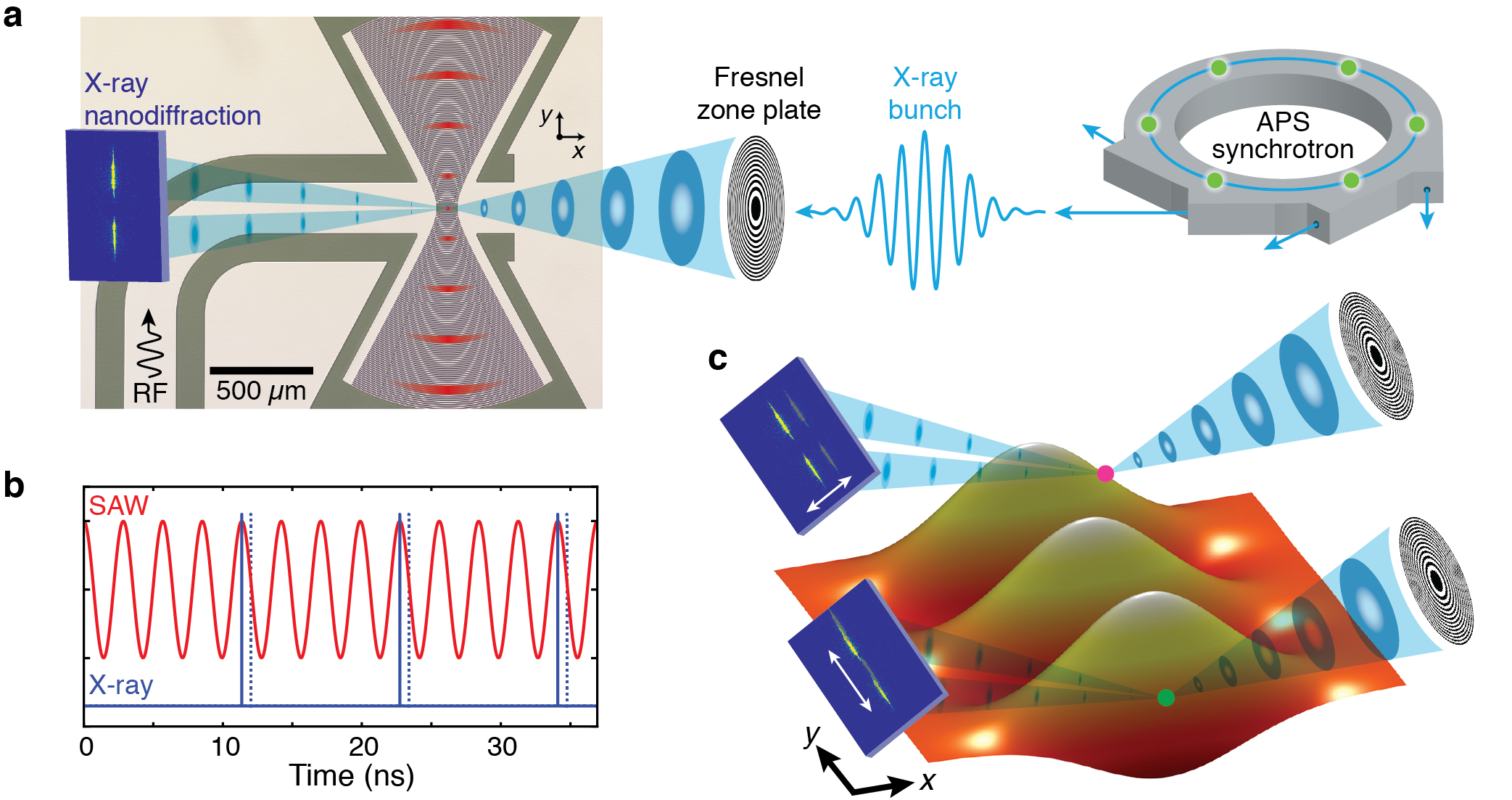}}
\caption{
\label{fig:schem} 
\textbf{Sketch of the stroboscopic Scanning X-ray Diffraction Microscopy (s-SXDM) experimental geometry.}  (\textbf{a}) Periodic bunch patterns of the Advanced Photon Source (APS) synchrotron are frequency matched to RF continuous wave excitation of a SAW (illustrated in red). Stroboscopic illumination allows for nanoscale SXDM of a virtually frozen wave. (\textbf{b}) The time domain surface displacement of the SAW (red curve - 1 nm peak-to-peak displacement at 352 MHz) is synchronized to the time structure of the X-ray illumination (blue curve - 22 ps rms width at approximately 88 MHz).  This allows a flexible measurement of the SAW amplitude by varying either time (detuned near-frequency match) or RF phase (at frequency match) to relatively displace the measurement to a new time slice (dotted blue line).  (\textbf{c}) The out-of-plane displacement of the SAW at a fixed point in time is represented by the orange isosurface, exhibiting both periodic lattice curvature along the wave longitudinal propagation direction \textit{y} (green dots) and transverse curvature along \textit{x} (magenta dots) induced by Gaussian focusing of the standing wave.  These curvatures induce orthogonal shifts in the far field diffraction pattern that oscillate as a function of the relative phase between the synchrotron time structure and the RF SAW excitation.  The two lobe feature of the nanodiffraction patterns is a consequence of the beam stop for the x-ray focusing optic (not shown). 
}
\end{figure*}

Surface acoustic waves (SAWs) have been previously observed with frequency-matched synchrotron X-ray scattering, historically with diffraction micro-topography \cite{Whatmore1982,Zolotoyabko2001}, more recently with iterative phase retrieval of surface topography utilizing SAW satellite peaks \cite{Reusch2013,Nicolas2014}, and also with photoemission electron microscopy techniques \cite{Foerster2017,Foerster2019}. These methods have typically utilized full field x-ray illumination with microscopic contrast mechanisms sensitive to a scalar projection of the lattice curvature at spatial resolutions down to 100 nm.  We report the development of an approach to imaging dynamic bulk crystal strain merging nano-focused hard X-ray scanning microscopy techniques with stroboscopic frequency synchronization at 3rd generation synchrotron sources enabling access to multiple vector strain components at 25 nm spatial resolution \cite{Holt2013}.  Using Bragg diffraction contrast as a scattering mechanism, our approach gives access to femtometer atomic displacements ($\Delta $c/c $\sim 10^{-5}$) in crystalline materials with a real-space, in-plane spatial resolution given by a beam focus on the order of tens of nanometers and at depths from tens to hundreds of microns.  By virtue of the high monochromaticity of the light, tuning the scattering specifically to the diffraction condition of the host material for quantum systems allows a strain contrast signal uncontaminated by the presence of capping material, transducer layers, or electrodes.  This ability to access far-from-surface lattice perturbations in complex systems is combined with an ultrafast pulsed X-ray illumination, which is determined by the bunch structure of the synchrotron storage ring \cite{Shenoy1988}.

Here, we investigate acoustic strain modulation using 22 ps rms X-ray pulses separated by 11.37 ns synchronized to a SAW excitation that penetrates microns into the surface.  In conventional pump-probe X-ray measurements, the synchrotron timing structure is used to synchronize an optical pump during pulsed stimulation of the sample \cite{Clark2013}.  Our measurement instead uses a frequency match of continuous radio frequency (RF) surface acoustic wave excitation to the ring time structure, in order to virtually slow or freeze periodic lattice fluctuations generated by a transducer fabricated on the 4H silicon carbide (SiC) host material without the need for a stimulation pulse.  We demonstrate the impact of this approach by correlating the dynamic lattice curvature measurement of driven lattice fluctuations with photoluminescence changes from point defects caused by acoustic driving and piezoelectric effects near an etched microscopic structure.

\section{Results}
\subsection{Gaussian Focusing of Surface Acoustic Waves in SiC}
Silicon carbide is a versatile and increasingly relevant material for quantum sensing and technological applications due to a rich variety of optically accessible defect spins with exceptionally long (millisecond) coherence times that are controllable up to room temperature  \cite{Weber2010,Koehl2011,Christle2014,Widmann2014}.  In comparison to diamond, SiC is available commercially on the wafer-scale with industrial growth processes established along with well understood micro and nanofabrication steps for the creation of electronic, mechanical, and optical functionality \cite{Bracher2017,Calusine2014}.  As robust as SiC neutral divacancy ground state spins are for storing quantum information, their excited state electronic energy levels can be manipulated and split with small amounts of crystalline strain ($< 10^{-6}$) in the host material \cite{Falk2014}.  SAW phonons have been used to demonstrate quantum manipulation of defect electronic orbitals in diamond \cite{Golter2016,Chen2018} and spin states in SiC \cite{Whiteley2018}.  This study demonstrates a local measurement of the lattice perturbations created by a SAW near a fabricated microscale structural defect in SiC, responsible for locally enhancing strain fluctuations around divacancies.

Propagating and stationary SAWs have widespread usage in RF signal processing and electronics applications \cite{Campbell1989} and are typically realized using an interdigitated transducer (IDT) on a piezoelectric crystal surface.  In this study we use a 500 nm thick piezoelectric transduction layer of sputtered aluminum nitride (AlN) on a 4H-SiC substrate with low film stress and fabricate an IDT to drive the SAWs.  The IDT contains a Ti under-metal layer for both improved metal-AlN adhesion and X-ray fluorescence mapping.  Considering that both 4H-SiC and AlN have isotropic Rayleigh wave velocities in their respective crystal planes, we apply Gaussian geometries, inspired by Gaussian optics and electromagnetism, to the IDT so as to form a nearly diffraction limited SAW spot size (Fig. \ref{fig:schem}a).  Gaussian focusing provides increased acoustic power near the focus, which is useful for increasing spin-phonon coupling in hybrid quantum systems while minimizing SAW diffraction losses in resonators \cite{Whiteley2018}.  The Gaussian IDT is geometrically designed to have a 1.25$\lambda$ spot size (SAW wavelength $\lambda$ = 19.03 $\mu$m) along \textit{x} and Guoy phase incorporated for the lowest order Hermite-Gauss mode. Additionally, the piezoelectric AlN film is etched away at the SAW beam waist where 3$\lambda$ of electrodes along the SAW propagation direction \textit{y} are removed to expose a window.  The purpose of the window is to remove extraneous microstructures in the 4H-SiC caused by the internal film stress in the AlN epilayer clamping the substrate surface.  Inside the IDT array, there is a standing wave produced by constructive interference from each of the individually phase matched electrodes.

\subsection{Stroboscopic X-ray Diffraction Microscopy}
The diffraction microscopy experiment sketched in Figure \ref{fig:schem} consists of 8 keV X-rays generated by the Advanced Photon Source focused to a 25 nm FWHM beam spot by an interlaced double Fresnel zone plate \cite{Comamala2011} at the Hard X-ray Nanoprobe Beamline operated by the Center for Nanoscale Materials, Argonne National Laboratory.  The sample is aligned to the 4H-SiC [0004] diffraction condition and the zone plate is raster scanned in real space relative to the sample position using an optomechanical nanopositioning system, allowing differential scanning of the nanofocused X-ray beam across the sample volume (Supplementary Figure 2) \cite{Winarski2012}.  The RF excitation for producing SAW phonons is matched to a multiple of the synchrotron storage ring frequency ($\sim 352$ MHz).  This creates Bragg diffraction from a virtually frozen wave curvature and strain that can be temporally swept by adjusting the relative phase of the RF signal generator to the synchrotron source using an RF IQ modulator (Fig. \ref{fig:schem}b).  In a second approach for phase sampling we add a small detuning to the SAW frequency relative to the ring frequency, which causes the signal to beat and evolve in phase at the detuning frequency. The resulting slow, time varying diffraction allows the SAW oscillation amplitude to be efficiently and periodically sampled without adjusting the electronics.  Both approaches yield similar results as a measure of local lattice distortion amplitude (Supplementary Figure 9). Here, the data we present are acquired in stroboscopic (time-sampled) mode with frequency detuning ($f_{ring}-f_{SAW} = 1~\pm~0.05$ Hz).  The far-field diffraction patterns are sensitive to strain components along the diffraction condition and lattice curvature or slope.  The two characteristic positions of high curvature on the Gaussian focused surface acoustic wave are marked as magenta and green dots in Fig. \ref{fig:schem}c. It is important to note that the directions of the lattice curvature gradient at the magenta and green points are mutually orthogonal, creating distinct oscillatory motions of the far-field diffraction patterns on the detector shown exaggerated in Fig. \ref{fig:schem}c.  These diffraction pattern oscillations are maximized at a characteristic set of positions where the curvature gradient reaches a maxima along the wave - either inflection points along the SAW propagation direction (longitudinal lattice slope - magenta dot) or gradient maxima of the Gaussian focusing (transverse lattice slope - green dot).  

\begin{figure*}[t]
\centerline{\includegraphics[width=170mm]{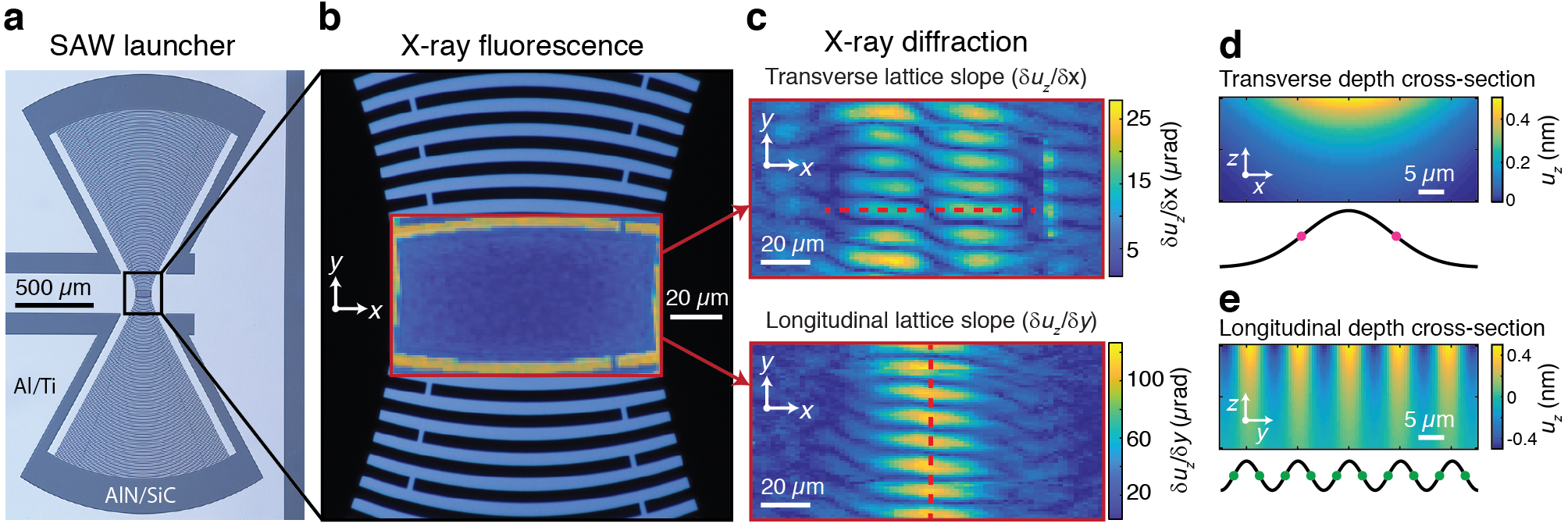}}
\caption{
\label{fig:SXDMSAWImaging} 
\textbf{Stroboscopic Scanning X-ray Diffraction Microscopy (s-SXDM) measurements made on a Gaussian SAW.}  (\textbf{a}) with the wave propagation direction aligned orthogonal to the \textit{x-z} scattering plane.  (\textbf{b}) The scanning X-ray Bragg diffraction measurements are registered relative to the patterned IDT using fluorescence from the Ti adhesion layer of the Al electrodes. The device background image is a dark field optical micrograph.  (\textbf{c}) The far-field X-ray diffraction pattern is sensitive to the local curvature (tilt) of the lattice planes calculated as the gradient of the displacement in two ordinal components.  The oscillatory transverse and longitudinal shifts of the X-ray diffraction pattern oscillation are independently detected as peak-to-peak displacements in the centroid position spot relative to the X-ray scattering plane.  (\textbf{d-e}) The expected depth dependence of an ideal Gaussian surface acoustic wave displacement ($u_z$) along the transverse \textit{x} (\textbf{d}) and longitudinal \textit{y} (\textbf{e}) directions at the dashed red lines in (c).  The transverse (longitudinal) curvature maxima, signified with magenta (green) dots, correlate with the double (single) lobed periodic features, respectively, experimentally observed in (c).  The calculated displacements correspond to the experimentally observed curvature amplitude.
}
\end{figure*}

Results of the stroboscopic Scanning X-ray Diffraction Microscopy (s-SXDM) measurement are shown in Figure \ref{fig:SXDMSAWImaging}. The IDT shown in Fig. \ref{fig:SXDMSAWImaging}a is aligned relative to the scanning directions using X-ray fluorescence (XRF) from the Ti under-metal layer of the Al patterned electrodes as in Fig. \ref{fig:SXDMSAWImaging}b.  At each scanning position twenty images of the far field diffraction pattern are acquired at a 50 millisecond detector exposure time, allowing for a stroboscopic sampling of one full oscillatory period - this permits efficient spatial mapping of amplitudes in time-sampled mode at a loss of the phase sensitivity possible with the IQ modulated mode.  Oscillatory motions of the diffraction patterns are then separated into two components, one lying within the diffraction plane (\textit{x}-axis) and one orthogonal to the diffraction plane (\textit{y}-axis).  These motions were primarily due to curvature inversion in the transverse and longitudinal (propagation) directions of the SAW, respectively.  The images in Fig. \ref{fig:SXDMSAWImaging}c are spatial maps of the amplitude of oscillation of the diffraction pattern during one full period - these are calculated from peak-to-peak dynamic X-ray diffraction centroid shifts relative to the mean value at each position.  This measure is used to remove spatially varying, time-independent lattice strain and curvature components in order to highlight SAW dynamics.  Disorder visible in the longitudinal picture is primarily due to complex internal reflections from the AlN film windowing visible near the edges of both maps in Fig. \ref{fig:SXDMSAWImaging}c.  Peak-to-peak displacement measurements fit to this curvature are consistent with a wave of $\sim$1 nm surface displacement amplitude over a 10 $\mu$m half-period (Fig.  \ref{fig:SXDMSAWImaging}d,e).  The curvature is sampled along a $\sim3$~$\mu$m X-ray extinction length expected at the [0004] reflection oriented at a scattering angle of $\approx18^{\circ}$, resulting in a 1 $\mu$m depth sampling of the transverse and longitudinal curvature fluctuations. No fluctuations are measured when the SAW power was reduced to zero (Supplementary Figure 8).

\subsection{Imaging Structural Defects}
\begin{figure}[t]
\centering
\includegraphics[width=89mm]{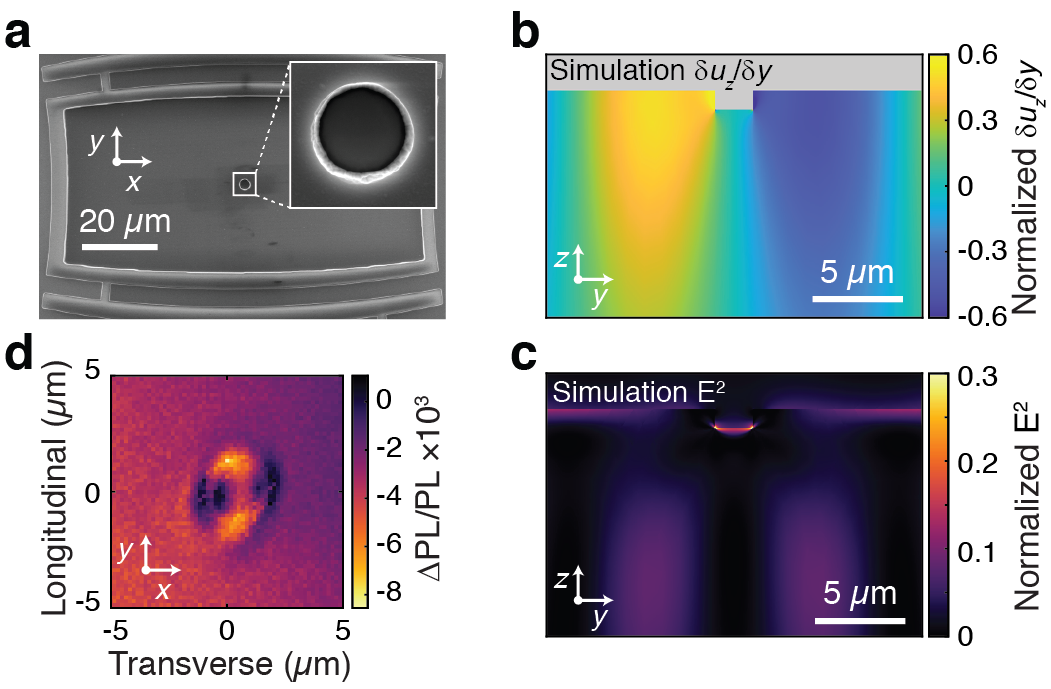}
\caption{
\label{fig:StructuralDefect} 
\textbf{Structural defect and internal piezoelectric response.} (\textbf{a}) SEM image of model structural defect, an etched pit, located at the center of the SAW beam waist. (\textbf{b,c}) Simulated depth cross-section of the longitudinal lattice curvature and electric field magnitude (E$^2$) in the SiC substrate.  The finite element models are computed using COMSOL Multiphysics and show a time slice during continuous RF excitation of the SAW. Although the etch pit is located at a node in longitudinal curvature, increased strain and piezoelectric effects happen at the pit's base and corners. (\textbf{d}) Enhanced dynamic piezoelectric effects at the pit are independently and all-optically mapped by native quantum defects using Electrometry by Optical Charge Conversion \cite{Wolfowicz2018}, which is sensitive to the local E$^2$. Peaks of signal are visible at the longitudinal pit edges, and the background is from stray electric fields (see Supplementary Figure 4 for depth dependence of PL contrast).
}
\end{figure}

Acoustic waves are often scattered by interactions with fabricated objects or lattice defects. In order to assess the divacancy response to locally induced curvature and strain, we fabricate a structural defect into the center of the SiC scanning window (Fig. \ref{fig:StructuralDefect}a) to perturb the Gaussian SAW.  This structural defect consists of a pit dry etched ($\sim$2.7 $\mu$m diameter, 1 $\mu$m deep) into the SiC surface.  The dynamic (time-dependent) strain perturbation of the etch pit in response to the acoustic wave is expected to be relatively large, as shown by a mechanical model in Fig. \ref{fig:StructuralDefect}b.  This effect results in locally enhanced strain and piezoelectric effects in the 4H-SiC when the SAW is present, as simulated in Fig. \ref{fig:StructuralDefect}c. Next, we probe the optical response of native divacancy defect ensembles, which emit photoluminescence (PL) in the near-infrared.  By employing the technique Electrometry by Optical Charge Conversion (EOCC) \cite{Wolfowicz2018}, we all-optically map the divacancy ensemble charge state population in order to probe dynamic electric fields induced piezoelectrically in the vicinity of the microscopic structural defect.  In this methodology, the optical charge conversion rates are sensitive to fluctuations in the local electric field (E$^2$), which we measure in the steady state by simultaneously illuminating the divacancy ensembles with both ultraviolet and near-infrared light.  We find that PL contrast during SAW excitation is spatially maximized near the structural defect (Fig. \ref{fig:StructuralDefect}d). These results correlate with the s-SXDM measurements of dynamic lattice curvatures where internal acoustic strains, detected by electric fields in the piezoelectric bulk 4H-SiC, are strongly enhanced near the etch pit corners. Furthermore, the depth dependence of PL contrast from the divacancies (Supplementary Figure 4c) indicates that the dynamic strain and electric fields are located microns away from the SiC surface.  This is in agreement with our numerical simulations (Fig. \ref{fig:StructuralDefect}c and see Supplementary Figure 7) that reveal the linear strain amplitude near the structural defect base is over three times greater compared to strain from the propagating SAW amplitude alone.   

\begin{figure*}[t]
\centerline{\includegraphics[width=170mm]{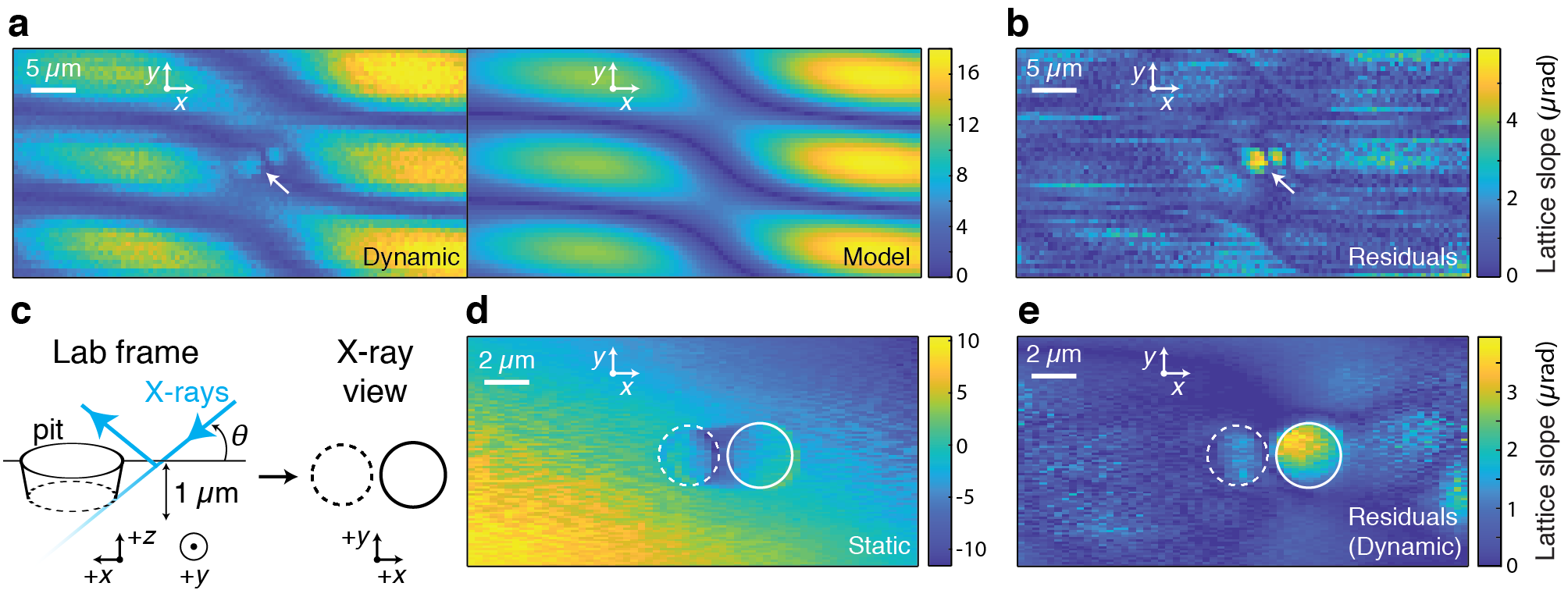}}
\caption{
\label{fig:NearDefectSXDM} \textbf{Near structural defect dynamic curvature measured via s-SXDM.} (\textbf{a}) Real-space map of the measured diffraction oscillation amplitude, a white arrow show presence of the fabricated pit in the SiC surface. The expected instantaneous curvature is predicted with a simple phenomenological model including Gaussian focusing and wave periodicity as free parameters.  The weak S-shaped connection of the bimodal transverse curvature originates from a small ($\sim 2~\deg$.) misalignment of the diffraction plane to the wave propagation direction.  (\textbf{b}) The best fit of the model parameters were then used to subtract the curvature of the unperturbed SAW, allowing visualization of a locally trapped high amplitude wave near the pit (arrow).  The spatial step size in (a,b) is 500 nm, with a 25 nm spot size. (\textbf{c}) SXDM sketch of the fabricated structural defect with a scattering angle $\theta \approx 18$~degrees. The surface coordinate scanning in the \textit{x-z} plane is corrected for sample angle relative to X-ray beam.  (\textbf{d}) Nanoscale imaging near the pit of static strain, or lattice curvature, from the mean \textit{x} centroid shift. (\textbf{e}) Dynamic transverse curvature residuals from the rms \textit{x} centroid shift during stroboscopic imaging with a theoretical SAW model subtracted. A solid and dashed circle represent the upper and lower edges of the pit for reference.  The spatial step size in (d,e) is 50 nm, with a 25 nm spot.
}
\end{figure*}

We directly measure the dynamic transverse curvature amplitudes near the etch pit by s-SXDM (Fig. \ref{fig:NearDefectSXDM}a).  The expected instantaneous curvature of an unperturbed SAW is predicted with a simple phenomenological model assuming only Hermite-Gauss modes from the IDT. We find that an additional rotation degree of freedom relative to the scattering plane explains the weak S-shaped connections of the bimodal transverse curvature.  This prediction is subtracted from the experimentally observed curvature amplitude in order to highlight the near-defect behavior, the results of which are shown in Fig. \ref{fig:NearDefectSXDM}b.  The etch pit geometry was chosen to be a fraction of a SAW wavelength, as small as possible for highest localized focusing from a single wave number while still accessible by optical spatial resolution limits.  This structure was intended primarily to create a local focusing of the oscillatory curvature which is most clearly visible as deviations from the phenomenological model in the transverse direction.  

Combining the nanometer-scale spatial resolution and stroboscopic feature of our X-ray imaging technique allows us to locally image both the static and actively driven lattice distortions around the etch pit.  Static local strain induced by the dry etch process is relatively small and nonetheless can be directly visualized by the mean (time independent) diffraction pattern recorded at each scanning position (Fig. \ref{fig:NearDefectSXDM}c,d). The static bound strain results can be understood as compressive (tensile) strain in the upper (lower) pit corners, respectively, which average and cancel each other when scanning through the center. At a higher degree of visibility in the dynamic variations (Fig. \ref{fig:NearDefectSXDM}e) compared to an outgoing wave, and trapped within the diameter of the etch pit, we observe a strong bimodal peak. The maps of longitudinal and transverse lattice curvatures, including SAW model for completeness, are shown in Supplementary Figure 10. By raster scanning the beam and correlating the resulting bound strain, we determine that this trapped acoustic feature is microns below the surface and varying in time, demonstrating the value of this synchrotron microscopy method as a local direct measurement of defect interactions with acoustic waves. Such an effect could in principle be used to locally focus acoustics near quantum relevant defects within a nanostructure in order to achieve stronger time-varying distortions for state manipulation.

\section{Discussion}
We demonstrated nano-focused, direct-space Bragg diffraction microscopy using the synchrotron timing structure to stroboscopically image time-varying lattice curvature.  This methodology provides both higher spatial resolution than previous acoustic wave imaging methods, and more importantly allows direct measurement of independent ordinal components of lattice curvature and strain \cite{Holt2014} rather than a single scalar projection of these quantities as in previous full field topographic and photoemission microscopy studies. This approach directly enables  the further development of nano-focused coherent diffraction imaging techniques such as 3D Bragg Projection Ptychography \cite{Hruszkewycz2017} which will now have the potential to image time-varying acoustic strain at sub-10~nm 3D resolution near far-from-surface lattice defects.

We imaged local dynamic lattice perturbations induced by a Gaussian focused SAW interacting with a fabricated structural defect in 4H-SiC and correlated these perturbations with divacancy defect optical measurements.  Our nano-focused scanning hard X-ray diffraction imaging approach, based on the frequency matching of a synchrotron X-ray pulse structure to a RF transducer, can be extended to real-space imaging of crystal strain in the time domain with full control over the acoustic phase. The induced strain and dynamic lattice fluctuations observed hold important consequences for nanomechanical system engineering with local quantum defects, as indicated by the EOCC photoluminescence contrast enhancement near the structural defect.  A key feature enabled by this methodology is removing ambiguity in assessing quantum structure-function relationships, through the congruent ability to image in real-space lattice fluctuations that directly relate to photoluminescence changes from optically-active point defects.  Beyond semiconductors, this method is generally applicable to acoustically manipulated structures and crystallographic defects where simultaneous picosecond time, nanoscale spatial, and sub-picometer strain displacement sensitivity can be used for unique local visualization of mechanical energy transduction.

\section{Methods}
\subsection{Sample Fabrication}
The high purity semi-insulating 4H-SiC substrate from Cree Inc. had $\sim$500 nm AlN sputtered by OEM Group Inc. onto the wafer Si-face with low film stress.  The IDT, comprising of 20 nm thick Ti and 150 nm thick Al, was fabricated on the AlN/SiC substrate. The circular pit (Fig. \ref{fig:StructuralDefect}a) was measured to be $\approx950$~nm deep by AFM using an Asylum Research Cypher S and the depth was also confirmed with laser confocal microscopy using an Olympus LEXT OLS5000.  All layers were processed by optical lithography techniques and inductively coupled plasma etching.  Extensive fabrication and device characterization details are shown in the Supplementary Information.

\section{Data Availability}
All data are available upon request to the corresponding author.

\section{Acknowledgements}
The fabrication of the surface acoustic wave devices was supported by the Air Force Office of Scientific Research.  The experimental design and surface acoustic wave microwave driving and synchronization was supported by the US Department of Energy, Office of Science, Basic Energy Sciences, Materials Sciences and Engineering Division.  The SXDM measurements were performed at the Hard X-ray Nanoprobe Beamline operated by the Center for Nanoscale Materials and the Advanced Photon Source, both Office of Science user facilities supported by the U.S. Department of Energy, Office of Science, Office of Basic Energy Sciences, under Contract No. DE-AC02-06CH11357.  This work made use of shared facilities supported by the NSF MRSEC Program under DMR-0820054 and the Pritzker Nanofabrication Facility of the Institute for Molecular Engineering at the University of Chicago, which receives support from Soft and Hybrid Nanotechnology Experimental (SHyNE) Resource (NSF ECCS-1542205), a node of the National Science Foundation’s National Nanotechnology Coordinated Infrastructure.  The authors thank Stephan O. Hruszkewycz, Paul C. Jerger, Brian B. Zhou and Paul G. Evans for careful reading of the manuscript.

\section{Author Contributions}
S.J.W. fabricated, simulated, and characterized the devices. F.J.H. and M.V.H. conceived the experiments and S.J.W., F.J.H., and M.V.H. carried out the X-ray diffraction imaging experiments. G.W. performed the optical measurements. D.D.A. and M.V.H. advised on all efforts. All authors contributed to discussions and production of the manuscript.

\section{Competing Interests}
The authors declare no competing interests.

\bibliographystyle{unsrt}
\bibliography{thebibliography}

\end{document}